\documentclass{aa}
\usepackage{times}
\usepackage{graphics}
\usepackage{latexsym}
\usepackage{epsf}
\begin{document}
%\topmargin0.0cm
%
%%%\thesaurus{06(03.20.1; 08.03.4; 08.09.2: CIT\,3; 08.13.2; 08.16.4; 13.09.6)}
%
\title{
A multi-wavelength study of the oxygen-rich AGB star \object{CIT\,3}:
Bispectrum speckle interferometry  and dust-shell modelling
%\thanks{Based on observations with  the SAO 6\,m telescope operated by the
%Special Astrophysical Observatory, Russia}
}
\author{
K.-H.\ Hofmann\inst{1}\and
Y.\ Balega\inst{2}\and
T.\ Bl\"ocker\inst{1}\and
G. Weigelt\inst{1}
}
\institute{
Max--Planck--Institut f\"ur Radioastronomie, Auf dem H\"ugel 69,
D--53121 Bonn, Germany \\
(hofmann@mpifr-bonn.mpg.de, bloecker@mpifr-bonn.mpg.de, 
 weigelt@mpifr-bonn.mpg.de)
\and
Special Astrophysical Observatory, Nizhnij Arkhyz, Zelenchuk region,
Karachai--Cherkesia, 35147, Russia (balega@sao.ru)
}
\offprints{K.-H.\ Hofmann}
%\mail{}
%

%%%\date{submitted: \today}
%%\date{this version: \today}
%%%\date{revised version: \today}
%%%\date{final version edp: \today}
%%%\date{received ~~~~ / accepted ~~~~}
\date{Received date /  accepted date}
\titlerunning{ 
The oxygen-rich AGB star \object{CIT\,3}:
Bispectrum speckle interferometry  and dust-shell modelling
}
\authorrunning{K.-H. Hofmann et al.}
\abstract{
CIT\,3 is an oxygen-rich
long-period variable evolving along the Asymptotic Giant Branch and
is one of the most extreme infrared AGB objects.
Due to substantial mass loss it is surrounded by an optically thick dust shell
which absorbs almost all visible light radiated by the star and
finally re-emits it in the infrared regime.
We present the first near infrared bispectrum speckle-interferometry
observations of  CIT\,3 in the $J$-, $H$-, and $K^{\prime}$-band.
The  $J$-, $H$-, and $K^{\prime}$-band resolution is 
48\,mas, 56\,mas, and 73\,mas, resp.
The interferograms were obtained with the Russian 6\,m telescope at the 
Special Astrophysical Observatory.
While CIT\,3 appears almost spherically symmetric in the
$H$- and $K^{\prime}$-band
it is clearly elongated in the $J$-band along a symmetry axis of position angle
$-28\degr$. Two structures  can be identified: a compact
elliptical core and a fainter north-western fan-like structure.
The eccentricity of the elliptical core, given by the ratio of
minor to major axis, is  approximately $\varepsilon$=123\,mas/154\,mas=0.8.
The full opening angle of the fan amounts to approximately $40\degr$.
%%%i.e.\ it extends from position angle $-8\degr$ to $-48\degr$.
%
Extensive radiative transfer calculations have been carried out 
and confronted with the observations taking into account 
the spectral energy distribution ranging from 1\,$\mu$m to 1\,mm, 
our near-infrared visibility functions at  1.24\,$\mu$m,
1.65\,$\mu$m and 2.12\,$\mu$m, as well as 11\,$\mu$m ISI interferometry.
The best model found to match the observations refers to a cool central star
with $T_{\rm eff}=2250$\,K which is surrounded by an optically thick dust shell with
$\tau (0.55\mu m) = 30$. 
The models give a central-star diameter of $\Theta_{\ast}=10.9$\,mas
and an inner dust shell diameter of  $\Theta_{1}=71.9$\,mas
being in line with lunar occultation observations.
The inner rim of the dust-shell is located at $r_{1}= 6.6 R_{\ast}$ and has a 
temperature of $T_{1}=900$\,K. The grain sizes were found to comply with 
a grain-size distribution according to Mathis et al.\ (1977) with 
 $n(a) \sim a^{-3.5}$, and  0.005\,$\mu {\rm m} \leq a  \leq 0.25\,\mu$m.
Uniform outflow models, i.e. density distributions with $\rho \sim 1/r^{2}$,
turned out to 
%%%match the near- to mid-infrared part of the SED and the 
%%%corresponding visibilities but to considerably 
underestimate the flux beyond 20\,$\mu$m. A two-component model existing of an inner 
uniform-outflow shell region ($\rho \sim 1/r^{2}$)
and an outer region where the density  declines more shallow as $\rho \sim 1/r^{1.5}$
proved to remove this flux deficiency and to give the best overall match of the
observations. The transition between both density distributions is at
$r_{2} = 20.5 r_{1}= 135.7 R_{\ast}$ where the dust-shell temperature has dropped
to $T_{2} = 163$\,K.
Provided the outflow velocity kept constant,
the more shallow density distribution in the outer shell
indicates  that mass-loss has decreased with time in the past of CIT\,3. 
Adopting $v_{\rm exp}=20$\,km/s, the termination of that mass-loss
decrease and 
the begin of the uniform-outflow phase 
took place 87\,yr ago. The present-day mass-loss rate can be determined to be 
$\dot{M} = (1.3-2.1) \cdot 10^{-5}$\,M$_{\odot}$/yr for $d=500-800$\,pc. 
\keywords{
Techniques: image processing ---
Circumstellar matter ---
Stars: individual: CIT\,3 ---
Stars: mass--loss ---
Stars: AGB and post-AGB ---
Infrared: stars
}
}
\maketitle
\section{Introduction}  \label{Sintro}
Within the framework of the Two Micron Sky Survey or Infrared Catalogue
(IRC, Neugebauer \& Leighton \cite{NeuLei69}),
Ulrich et al.\ (\cite{UlrEtal66}) published a list of
14 bright (at 2.2\,$\mu$m) and very red objects being of potential interest
for the study of cool and dust-enshrouded stars.
Among these so-called CIT objects is
\object{CIT\,3}  (= WX\,Psc = IRC\,+10\,011 = IRAS\,01037+1219), an oxygen-rich
long-period variable evolving along the Asymptotic Giant Branch (AGB)
and one of the most extreme infrared AGB objects.

\object{CIT\,3} belongs to the very late-type AGB stars having a
spectral type of M9-10 (Dyck et al.\ \cite{DyckEtal74},
Lockwood \cite{Lock85}). It is very faint in the optical
($V > 20$\,mag) but bright in the infrared ($K$=2\,mag).
Given this spectral type and the exceptionally large color index of
$V-K \ga 18$, its effective temperature can be estimated to be
$\simeq 2500$\,K or less considering the effective temperature scales of 
Dyck et al.\ (\cite{DyckEtal96},\cite{DyckEtal98}) and 
Perrin et al.\ (\cite{PerEtal98}).
\object{CIT\,3} is a regular pulsational variable with a mean infrared
variability period of 660 days (Harvey et al.\ \cite{HarEtal74},
Le Bertre \cite{LeB93}, Whitelock et al.\ \cite{WhiEtal94}).
The stellar surface of \object{CIT\,3} is eroded by strong stellar winds
with current mass-loss rates close to $10^{-5}$\,M$_{\odot}$/yr. For example,
various CO observations indicate rates in the range of 
%%%0.79, 0.85, 1.4, 2.4 and 2.6 x $10^{-5}$\,M$_{\odot}$/yr, resp.
0.79 up to  2.6 x $10^{-5}$\,M$_{\odot}$/yr, resp.
(Neri et a.\  \cite{NeriEtal98},  Loup et al.\ \cite{LoupEtal93},
 Knapp \& Morris \cite{KnaMor85}, Sopka et al.\ \cite{SopEtal89},
 Knapp et al. \cite{KnaEtal82}).
Accordingly, \object{CIT\,3} is surrounded by a thick dust shell.
This dust shell absorbs almost all visible light radiated by the star and
finally re-emits it in the infrared regime.
The circumstellar envelope of \object{CIT\,3} shows
SiO (Dickinson et al.\ \cite{DickEtal78}, Desmurs et al.\ \cite{DesEtal2000}),
H$_{2}$O (Dickinson \cite{Dick76}, Bowers \& Hagen \cite{BowHag84}) and
OH maser emission (Wilson et al.\ \cite{WilEtal70},
Olnon et al.\ \cite{OlnEtal80}). The expansion
velocity of the circumstellar shell amounts to 19\,km/s
if measured in the  1612\,MHz OH maser lines
(Baud \cite{Baud81}), and 18\,km/s to 23\,km/s, with a mean value of 21\,km/s,
if measured in CO(1-0) and CO(2-1) transitions
(Loup et al.\ \cite{LoupEtal93}).
%%%%%%%{\tt high-lattitude star, scale height) \\

The angular radius of the OH emitting shell region has been determined
by Very Large Array (VLA) measurements
to be 4\farcs0 (Baud \cite{Baud81}) and 4\farcs4 (Bowers et al.
\cite{BowEtal83}), resp. The corresponding linear OH radius can directly be
determined by the so-called phase-lag method utilizing the 
phase difference between the red-shifted and blue-shifted peak
in the 1612Mhz OH spectrum during the regular IR and OH light variations. 
Due to the difference in light travel time between the emission from the front
and the back of the shell, the variations of these two masing regions
are out of phase at Earth. Jewell et al.\ (\cite{JewEtal80}) and
van Langevelde et al.\ (\cite{vanLEtal90}) observed a phase lag of
25.5\,d and 34\,d, resp., resulting in a linear OH shell radius of
$R_{\rm OH} = 3.3\cdot 10^{14}$cm and $R_{\rm OH} = 4.4\cdot 10^{14}$cm, resp.
Combining the VLA observations with the phase-lag measurements provides
an independent method for determining the distance to \object{CIT\,3}.
For instance, Baud (\cite{Baud81}) derived $d=570$\,pc using the phase lag of
Jewell et al.\ (\cite{JewEtal80}), the phase lag of 
van Langevelde et al.\ (\cite{vanLEtal90}) gives $d=740$\,pc.
The somewhat larger angular OH radius of Bowers et al.\ (\cite{BowEtal83})
result in correspondingly smaller distances, i.e. 520\,pc and 670\,pc. 
These direct determinations can be compared with other, indirect, methods.
For instance, Hyland et al.\ (\cite{HylEtal72}) give a distance of 510\,pc
based on the measured flux at maximum light
($1.1\cdot 10^{-9}$Wm$^{-1}$) and on the assumption that the
maximum stellar luminosity amounts to $10^{4} L_{\odot}$.
Adopting $L=2 \cdot 10^{4} L_{\odot}$, would result into 720\,pc.
Whitelock et al.\ (\cite{WhiEtal94}) applied period-luminosity relations
to obtain the absolute bolometric magnitude which, together with the
apparent magnitude, finally leads to  $d=820$\,pc. In summary, the distance
to \object{CIT\,3} appears to be in the narrow range from 500 to 800\,pc.

Zappala et al.\ (\cite{ZapEtal74}) measured the size of \object{CIT\,3}
at 2.2\,$\mu$m,  10\,$\mu$m, and  20\,$\mu$m by means of lunar occultations.
They fitted the disappearence curves at 2.2\,$\mu$m and  10\,$\mu$m
with two-disks fits. At 2.2\,$\mu$m, they obtained a small component with
an angular diameter of 13\,mas contributing 65\% of the flux
and a larger one with an diameter of 65\,mas contributing 35\% of the flux.
The smaller component was interpreted to be the central star whereas the
larger one represents the inner region of the dust shell. At  10\,$\mu$m,
the fit yielded one component of approximately 65\,mas diameter providing
15\% of the flux,  and one of  135\,mas providing 85\% of the flux.
In summary, \object{CIT\,3} appeared to consist of
a central star with a diameter of 13\,mas, surrounded by a dust shell
with typical diameters of 65\,mas at  2.2\,$\mu$m and 135\,mas at 
10\,$\mu$m. 

Recent mid-infrared visibility measurements were carried out by
Sudol et al.\ (\cite{SudEtal99}) and
Lipman et al.\ (\cite{LipEtal2000}).
Sudol et al.\ (\cite{SudEtal99}) quote an upper limit of 0\farcs63
for the 11\,$\mu$m shell diameter since 
\object{CIT\,3} was not resolved in their 2.3\,m telescope images.
Interferometric measurements of Lipman et al.\ (\cite{LipEtal2000})
with the Infrared Spatial Interferometer (ISI) clearly resolve the
object at  11\,$\mu$m indicating an inner dust-shell diameter of 66\,mas
by means of corresponding radiative transfer models. 

There are several pieces of evidence that the dust shell of \object{CIT\,3}
deviates from spherical symmetry. 
Marengo et al.\ (\cite{MarEtal99}) reported evidence for an elongation
at 8.55\,$\mu$m based on 1.5\,m telescope data.
Neri et al. (\cite{NeriEtal98}) fitted their CO observations with two-component
Gaussian visibility profiles consisting of a circular and an elliptical
component. The circular flux-dominant component corresponds to a spherical
envelope with a 29\farcs6 diameter, the elliptical one to an axisymmetric
envelope structure with major and minor axis of 9\farcs8 and 6\farcs8, resp,
and a position angle of $-45\degr$ (E to N). 
%Marengo et al.\ (\cite{MarEtal99})
%presented mid-infrared imaging with a 1.5\,m telescope
%in the wavelength range from
%8.55\,$\mu$m to  16.64\,$\mu$m. At 8.55\,$\mu$m, \object{CIT\,3} appeared to be
%elongated with a major and minor axis of 3\farcs0 and 2\farcs3, resp. The
%position angle is rotated by additional $59\degr$ compared to the CO
%observations, and \object{CIT\,3} seems to be extended in N-S direction while
%being unresolved along the E-W axis. 

Finally, near-infrared polarimetry (Kruszewski \& Coyne \cite{KruCoy76},
McCall \& Hough \cite{McCHou80}) revealed that \object{CIT\,3} is
highly polarized in the $I$ and $J$-band with polarization degrees
of up to 8-9\%. The degree of polarization depends also on the phase.
In the $H$- and $K^{\prime}$-band polarization patterns are less dominant
(degree of polarization of the order of 1\%). The high degree of polarization
either indicates an alignment of the dust particles or an asymmetrical
dust-shell structure. 

In this paper we present the first bispectrum speckle-interferometry
observations of  \object{CIT\,3} at 1.24\,$\mu$m ($J$-band),
1.65\,$\mu$m ($H$-band), and 2.12\,$\mu$m ($K^{\prime}$-band).
In the $K^{\prime}$- and $H$-band,  the reconstructed images are diffraction-limited,
the resolution in the $J$-band is 88\% of the diffraction
limit. The  $J$-, $H$-, and $K^{\prime}$-band resolution amounts to  
48\,mas, 56\,mas, and 73\,mas, resp.
Radiative transfer calculations have been conducted to
model the spectral energy distribution as well as the various 
visibility functions in the near and mid-infrared. This paper is
organized as follows: In Sect.~\ref{Sobs} the bispectrum
speckle interferometric observations and data reduction are described.
Sect.~\ref{Smidir} resumes the results from available mid-infrared 
long-baseline interferometry. 
Sect.~\ref{Sphoto} summarizes the available photometry, spectrophotometry
and spectroscopy, and 
in Sect.~\ref{Sdusty}
radiative transfer modelling and its confrontation with the
observations is discussed. Finally, a summary and concluding remarks are given
in Sect.~\ref{Ssum}
\section{Observations and data reduction} \label{Sobs}
The \object{CIT 3} speckle interferograms
were obtained with the Russian 6\,m telescope at the Special Astrophysical
Observatory on September 19, 22 and 28, 1999.
These dates correspond to the phases $\phi=0.74$, 0.75 and 0.76 of the periodic
variability cycle of 660 days (see Sect.~\ref{SSvar}).
The speckle interferograms were
recorded with our HAWAII speckle camera (HgCdTe array
of 512$^2$ pixels, sensitivity from 1 to 2.5\,$\mu$m)
through interference filters with center wavelength/bandwidth of
1.24\,$\mu$m/0.14\,$\mu$m ($J$-band), 1.65\,$\mu$m/0.32\,$\mu$m ($H$-band) and
2.12\,$\mu$m/0.21\,$\mu$m ($K^{\prime}$-band).
Speckle interferograms of the unresolved stars HIP 7740 and HIP 6363 were
taken for the compensation of the speckle interferometry transfer function.
The observational parameters were as follows:
$J$-band observation (Sept.\ 28):
exposure time/frame 80~ms, number of frames 1963
(1268 of \object{CIT 3} and 695 of \object{HIP 6363});
$H$-band observation (Sept.\ 28):
exposure time/frame 20~ms, number of frames 2594
(1723 of \object{CIT 3} and 871 of \object{HIP 6363});
$K^{\prime}$-band observation (Sept.\ 19+22) :
exposure time/frame 25~ms, number of frames 6496
(3016 of \object{CIT 3} and 3480 of \object{HIP 7740});
1.24\,$\mu$m-seeing (FWHM) $\sim$1\farcs8;
1.65\,$\mu$m-seeing (FWHM) $\sim$1\farcs2;
2.12\,$\mu$m-seeing (FWHM) $\sim$1\farcs3;
$J$-band pixel size 13.4\,mas,
$H$-band pixel size 20.1\,mas,
$K^{\prime}$-band pixel size 26.4\,mas;
field of view 3\farcs4 ($K^{\prime}$-band) and 5\farcs1 ($H$-, $J$-band).  
%%%% diffraction-limit J: 43mas.
Images of \object{CIT 3} with 48\,mas ($J$-band),  56\,mas ($H$-band),
and 73\,mas ($K^{\prime}$-band) resolution were reconstructed from 
the speckle interferograms using the  bispectrum speckle-interferometry method
(Weigelt \cite{Wei77}, Lohmann et al.\ \cite{LohWeiWir83},
Hofmann \& Weigelt \cite{HofWei86}).
The $H$- and $K^{\prime}$-band images are diffraction-limited,
the resolution in the $J$-band is only slightly (12\%) below the
diffraction limit. The modulus of the object Fourier transform
(visibility function) was determined  with the speckle interferometry method
(Labeyrie \cite{Lab70}).
%%%%%%%%%%%%%%%%%%%%%%%%%%%%%%%%%%%%%%%%%%%%%%%%%%%%%%%%%%%%%%%%%%%%%%%%%%%%%
%%%% Images (top) and contour plots(bottom)  J,H,K
%%%%%%%%%%%%%%%%%%%%%%%%%%%%%%%%%%%%%%%%%%%%%%%%%%%%%%%%%%%%%%%%%%%%%%%%%%%%%
\begin{figure*}
\hbox{
\epsfxsize=5.5cm
%%%\mbox{\epsffile[-20 -20 226 226]{./plots/J-Band/CIT3-J4d.ima.eps}} \hfill
\mbox{\epsffile[-20 -20 226 226]{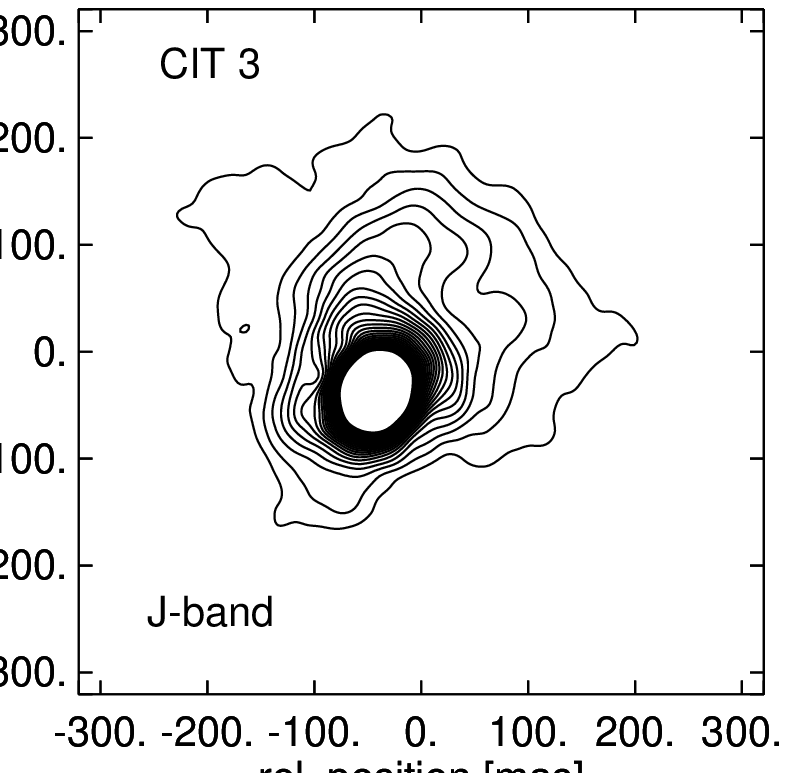}} \hfill
\epsfxsize=5.5cm
%%%\mbox{\epsffile[-20 -20 226 226]{./plots/H-Band/CIT3-H.ima.eps}}
\mbox{\epsffile[-20 -20 226 226]{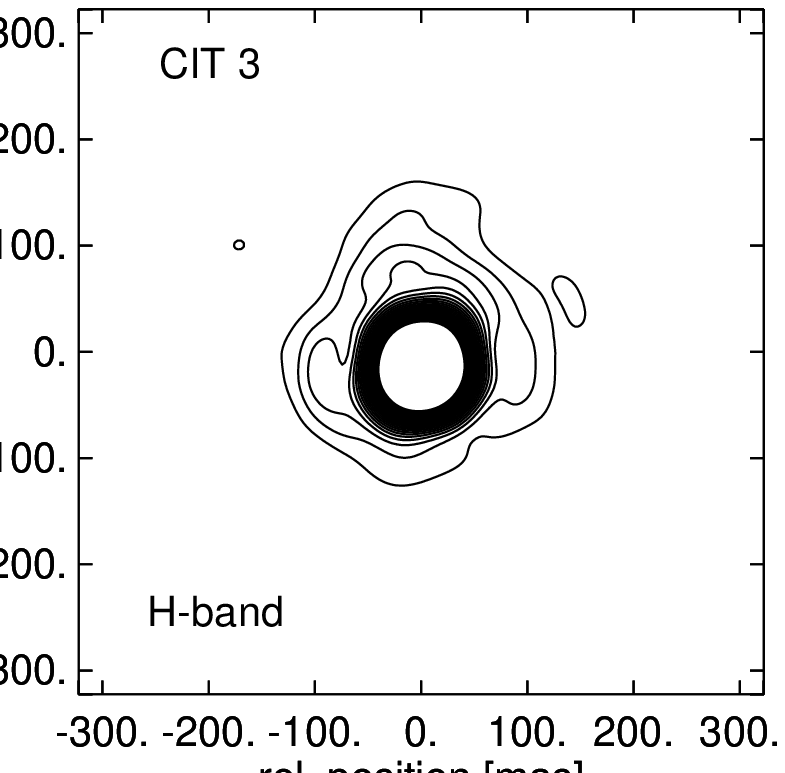}}
 \hfill 
\epsfxsize=5.5cm
%%%\mbox{\epsffile[-20 -20 226 226]{./plots/K-Band/CIT3-K.ima.eps}}
\mbox{\epsffile[-20 -20 226 226]{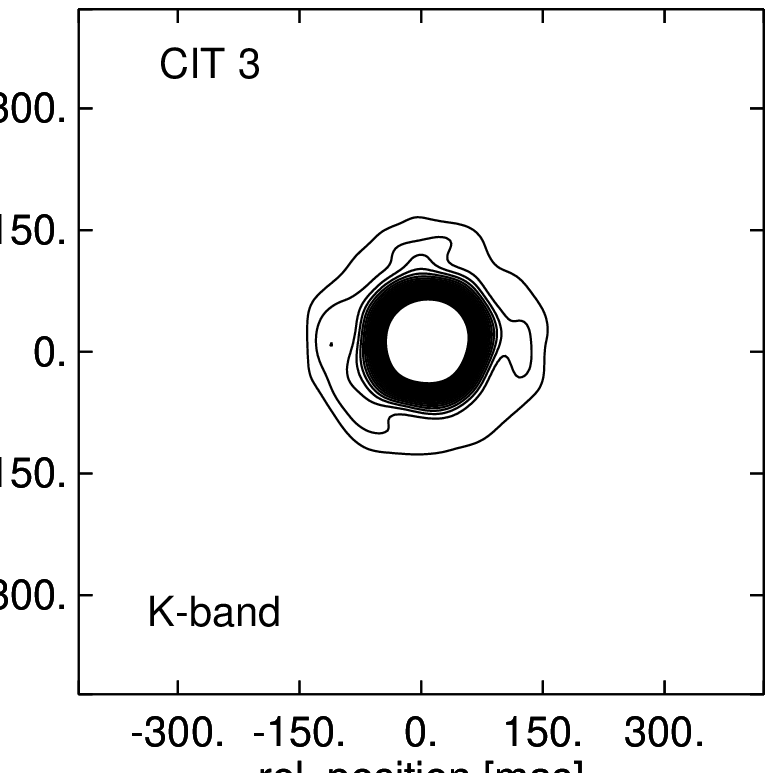}}
}
\caption{
Contour plots of the reconstructed \object{CIT\,3} images
at 1.24\,$\mu$m ($J$-band),  1.65\,$\mu$m ($H$-band), and 2.12\,$\mu$m
($K^{\prime}$-band) from left to right.
The  $J$-, $H$-, and $K^{\prime}$-band 
resolution amounts to  48\,mas, 57\,mas, and 73\,mas, resp. The images
in the $K^{\prime}$- and $H$-band  are diffraction-limited, the resolution of the
$J$-image is 88\% of the diffraction limit .
Contour levels are plotted from 1.5\% to 29.5\% of peak 
intensity in steps of 1\%. North is up and east is to the left.
}
\label{Fima} 
\end{figure*}
%%%%%%%%%%%%%%%%%%%%%%%%%%%%%%%%%%%%%%%%%%%%%%%%%%%%%%%%%%%%%%%%%%%%%%%%%%%%%
%%%% Visi (azimuth) (top) and Visi (cuts)  J,H,K
%%%%%%%%%%%%%%%%%%%%%%%%%%%%%%%%%%%%%%%%%%%%%%%%%%%%%%%%%%%%%%%%%%%%%%%%%%%%%
\begin{figure*}
\hbox{
\epsfxsize=5.75cm
%%%\mbox{\epsffile{./plots/J-Band/CIT3-J.vis.cut18.eps}} 
\mbox{\epsffile{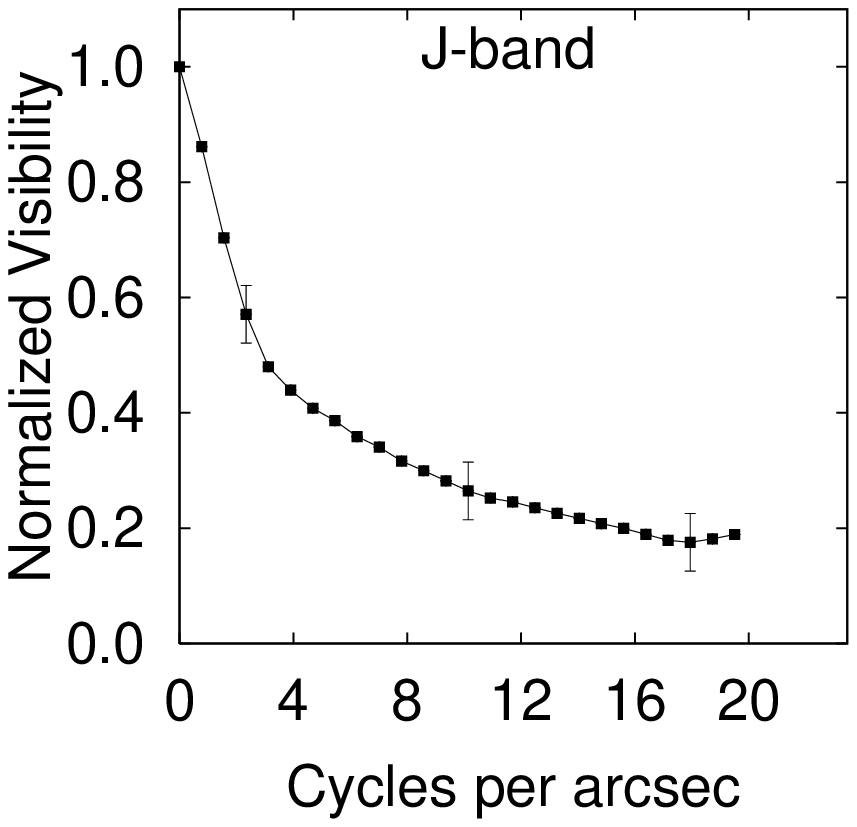}} 
\hspace*{0.25cm}
\epsfxsize=5.75cm
%%%\mbox{\epsffile{./plots/H-Band/CIT3-H.vis.eps}} 
\mbox{\epsffile{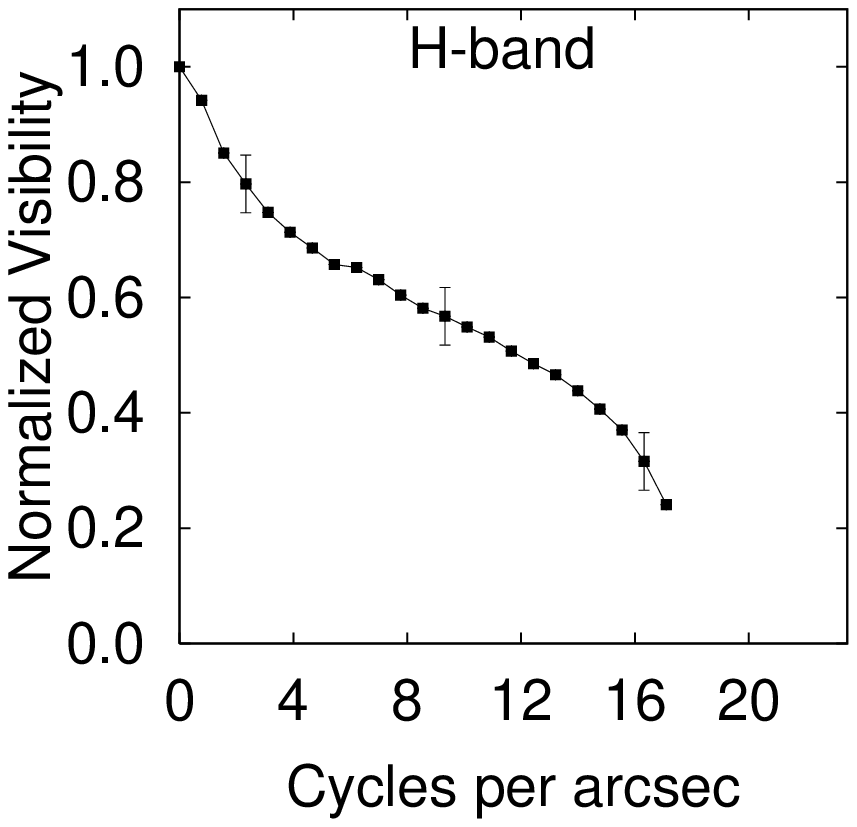}} 
\hspace*{0.25cm}
\epsfxsize=5.75cm
%%%\mbox{\epsffile{./plots/K-Band/CIT3-K.vis.eps}} 
\mbox{\epsffile{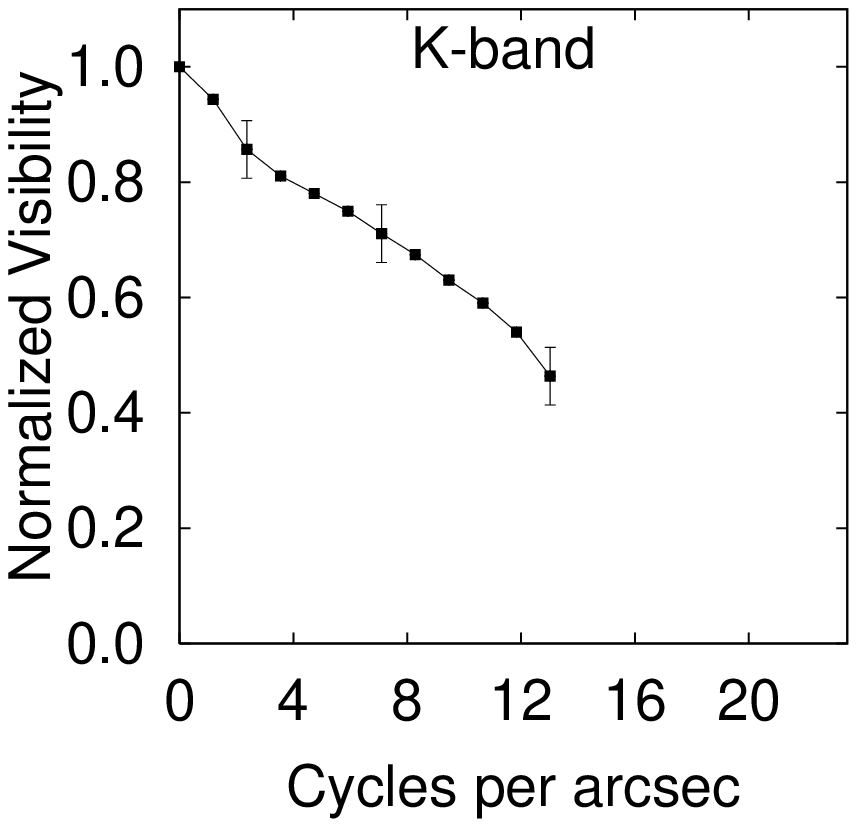}} 
}
\hbox{
\epsfxsize=5.75cm
%%%\mbox{\epsffile{./plots/J-Band/CIT3-J-Band.Schnitte.cut18.eps}} 
\mbox{\epsffile{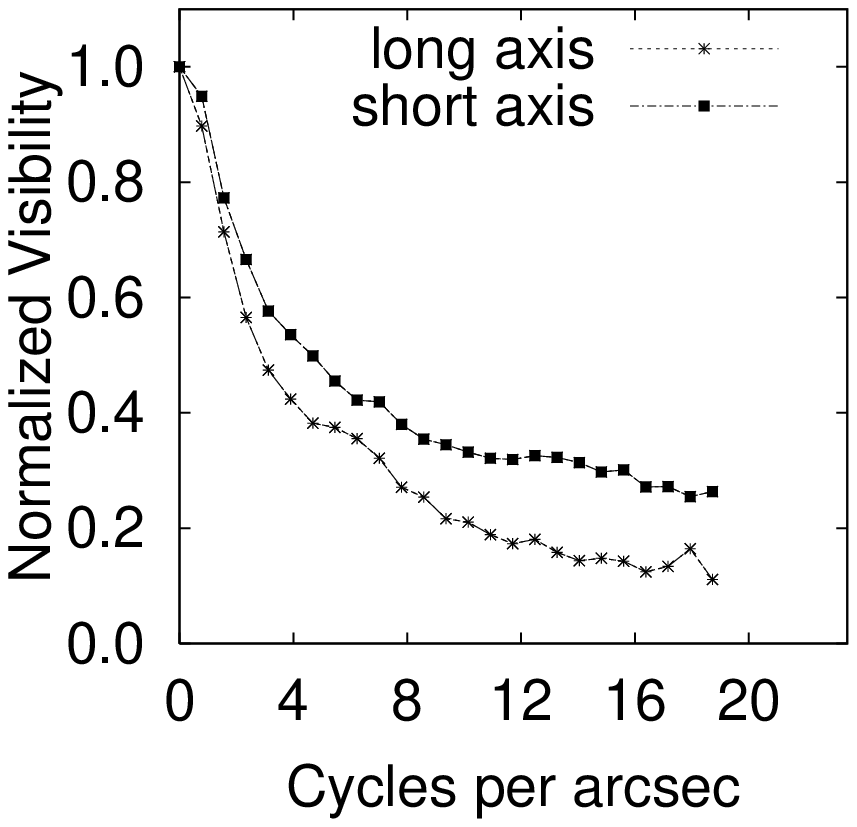}} 
\hspace*{0.25cm}
\epsfxsize=5.75cm
%%%\mbox{\epsffile{./plots/H-Band/CIT3-H.vis.schnitte.eps}} 
\mbox{\epsffile{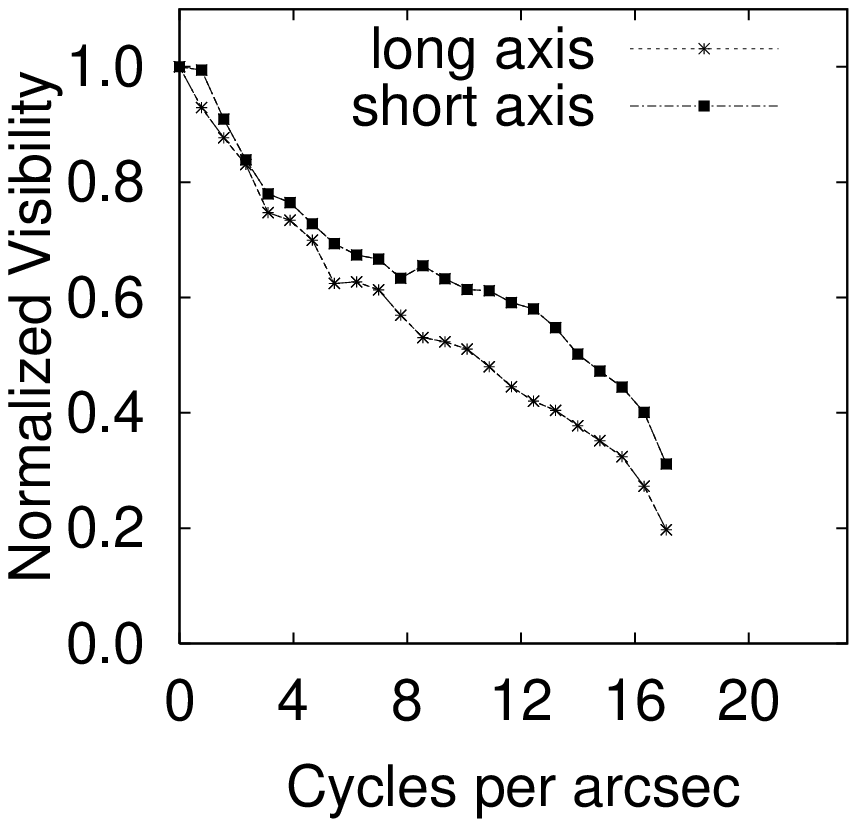}} 
\hspace*{0.25cm}
\epsfxsize=5.75cm
%%%\mbox{\epsffile{./plots/K-Band/CIT3-K.vis.schnitte.eps}} 
\mbox{\epsffile{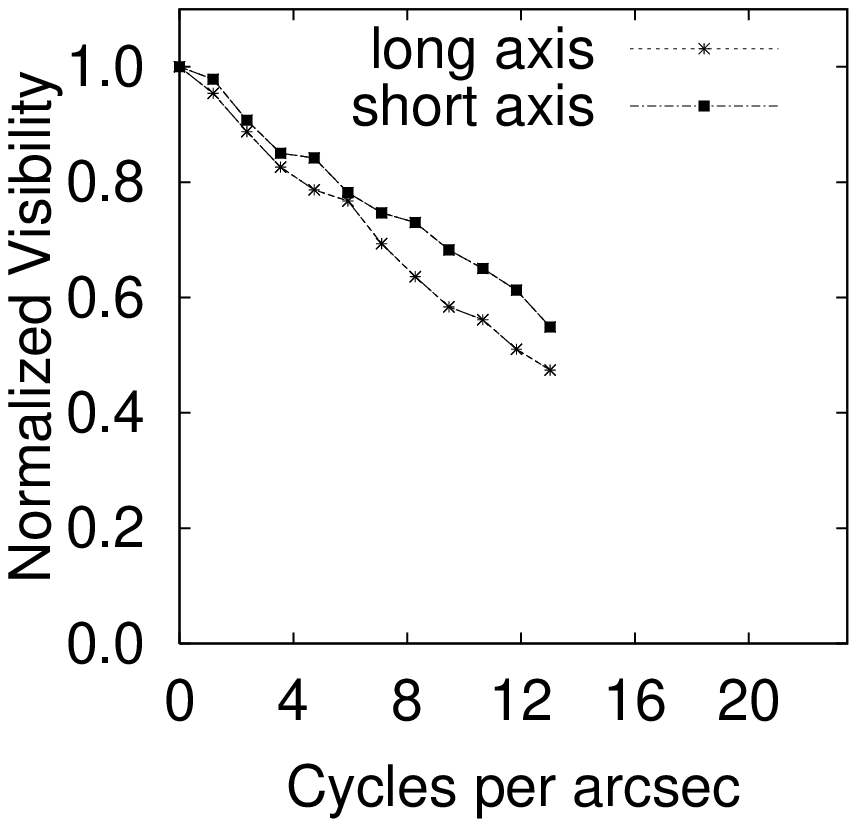}} 
}
\caption{
Visibility functions of \object{CIT\,3}
at 1.24\,$\mu$m ($J$-band),  1.65\,$\mu$m ($H$-band), and 2.12\,$\mu$m
($K^{\prime}$-band) from left to right.
The top panels show the azimuthally averaged visibilities whereas
the bottom panels display visibility cuts  
at position angles of $-28\degr$ (long axis) and $62\degr$ (short axis),
i.e.\ along and  perpendicular to the main symmetry axis of the elongated
compact structure of \object{CIT\,3} in the $J$-band (see Fig.~\ref{Fima}).
}\label{Fvis} 
\end{figure*}
%%%%%%%%%%%%%%%%%%%%%%%%%%%%%%%%%%%%%%%%%%%%%%%%%%%%%%%%%%%%%%%%%%%%%%%%%%%%%

Figure~\ref{Fima} shows the contour plots of the reconstructed 
\object{CIT\,3} images
at 1.24\,$\mu$m,  1.65\,$\mu$m, and 2.12\,$\mu$m.
Contour levels are plotted from 1.5\% to 29.5\% of peak intensity
in steps of 1\%. 
The $J$-band image is clearly elongated along a symmetry axis of position angle
$-28\degr$. More precisely, two structures  can be identified: a compact
elliptical core and a fainter north-western fan-like structure.
The eccentricity of the elliptical core, given by the ratio of
minor to major axis, is approximately  $\varepsilon$=123\,mas/154\,mas=0.8.
The full opening angle of the fan amounts to approximately $40\degr$,
i.e.\ it extends from position angles $-8\degr$ to $-48\degr$.
In the $H$-band the elongation of the core is hardly visible, and 
the fan structure is much weaker.
%The fan 
%appears to be smeared out, its opening angle increases by roughly
%$10\degr$, and the intensity valley between the  fan edges becomes flatter.
Finally in the $K^{\prime}$-band, the fan structure has almost disappeared being
only present on the 2.5\% intensity level, and \object{CIT\,3} looks almost
spherically symmetric. 

Figure~\ref{Fvis} displays the corresponding visibility functions of
\object{CIT\,3} at 1.24\,$\mu$m,  1.65\,$\mu$m, and 2.12\,$\mu$m, i.e.\
the azimuthally averaged visibilities as well as  visibility cuts  
at position angles of $-28\degr$ and $62\degr$,
i.e.\ along and  perpendicular to the main symmetry axis of the elongated
compact structure and the north-western outflow of \object{CIT\,3}
in the $J$-band (see Fig.~\ref{Fima}).
Towards the diffraction limit,
the azimuthally averaged visibility functions decline to $\sim 0.2$, almost
reaching a plateau, at  1.24\,$\mu$m, to  $\sim 0.25$ at  1.65\,$\mu$m and
to  $\sim 0.45$ at 2.12\,$\mu$m. In principle, but most obvious in the
$J$-band, all visibilities show the
same behaviour: a sharp decline at low spatial frequencies
($\la$ 3-4 cycles/arcsec) and an extended run of the curve of smaller slope
at larger spatial frequencies almost up to the diffraction limit.
This indicates the existence of a compact and of
an extended dust-shell component.
%The triangle shape of the visibility functions at low spatial frequencies
%points to ring-like intensity distributions being typical
%for optically thin dust-shells (see e.g.\
%Bl\"ocker et al.\ \cite{BloeEtal99}).

The visibility cuts along and perpendicular to the main
symmetry axis show similar features.
Additionally, as expected, the visiblity cuts along the symmetry axis decline
stronger than those perpendicular to it, 
evidencing a larger object extension in this direction. The typical differences
between the visibility cuts at large spatial frequencies amount up to
$0.15$ at 1.24\,$\mu$m and  1.65\,$\mu$m and
$\la 0.1$ at 2.12\,$\mu$m. In the $J$- and $H$-band this position angle
dependence can be considered as significant, whereas in the $K^{\prime}$-band
the differences are of the order of the error bar and thus are negligible. 
\section{Mid-infrared visibility} \label{Smidir}
Mid-infrared visibilities provide additional pieces of spatial information
since they are sensitive to the location of the outer, cooler regions of the
dust shell governed by thermal dust-emission, whereas near-infrared
visibilities are indicative for the inner, hot dust-formation zones and are
subject to scattering. 
The interferometric $11.15 \mu$m observations of
Lipman et al.\ (\cite{LipEtal2000}) with the Infrared Spatial Interferometer
at baselines of 4m, 9.6m and 16m are shown in Fig.~\ref{Fbestmod}.
Note, that
the data is not coeval but refers to variability phases (see Sect.~\ref{SSvar})
of 0.58-0.66 (4m), 0.71-0.76 (9.6m) and 0.66-0.71 (16m). 
%%%%%%%%%%%%%%%%%%%%%%%%%%%%%%%%%%%%%%%%%%%%%%%%%%%%%%%%%%%%%%%%%%%%%%%%%%%%%
%\begin{figure}[htbp]
%\epsfxsize=5.2cm
%\mbox{\epsffile{./plots/obs/aa_visi_11mu_data.ps}}
%\hfill
%\parbox[b]{2.9cm}{
%\caption[visi11]
%{Mid-infrared visibility functions of \object{CIT\,3}
% at $11 \mu$m according to the ISI observations at  4m, 9.6m and 16m baseline
% of Lipman et al.\ (\cite{LipEtal2000}) \vspace*{3ex}.
%}  \label{Fvis11}
%}
%\end{figure}
%%%%%%%%%%%%%%%%%%%%%%%%%%%%%%%%%%%%%%%%%%%%%%%%%%%%%%%%%%%%%%%%%%%%%%%%%%
%
\section{Spectral energy distribution} \label{Sphoto}
\subsection{Variability}   \label{SSvar}
\object{CIT\,3} is a regular variable with a mean infrared period of
660\,d. Infrared lightcurves are given in 
Harvey et al.\ (\cite{HarEtal74}), Le Bertre (\cite{LeB93}), and
Whitelock et al.\ (\cite{WhiEtal94}),  and radio lightcurves in
Herman \& Habing (\cite{HerHab85}). The period appears to be rather stable
with time and has not change since the 1970's. It depends slightly on the
wavelength, being somewhat shorter at longer wavelengths.
For instance,  Le Bertre (\cite{LeB93}) determined $P=668$, 661 and
654\,d at $\lambda=1.24$\,$\mu$m, 2.19\,$\mu$m, and 4.64\,$\mu$m, resp., and
Herman \& Habing (\cite{HerHab85}) found $P=632$\,d at $\lambda=18$\,cm.
The amplitudes of the brighness variations show, however, a strong
wavelength dependence in the near-infrared.  Le Bertre (\cite{LeB93}) gives
$\Delta m= 2.83$, 1.79, and 1.29 at 1.24\,$\mu$m, 2.19\,$\mu$m, and
4.64\,$\mu$m, resp. Then, at longer wavelenghts the amplitude seems to stay
roughly constant. Harvey found $\Delta m= 1.5$ at 10\,$\mu$m and
Herman \& Habing (\cite{HerHab85})  $\Delta m= 1.2$ at 18\,cm.

The spectrum of \object{CIT\,3}
shows a 9.7\,$\mu$m silicate feature in
partial self-absorption (see Fig.~\ref{Fbestmod}).
Monnier et al.\ (\cite{MonEtal98})
presented multi-epoch 8-13\,$\mu$m spectrophotometry (1994-1997)
covering more than one variability cycle, and 
showed that the spectral shape of the silicate feature
changes during the cycle. The
%%%becoming more narrow near maximum light. The
brightness variation between minimum and maximum light was $\sim 1.2$\,mag.

Thus, the variability-phase dependence
of the spectral energy distribution (SED) can be
characterized by wavelength independent variations of  $\Delta m \sim 1.2$
for $\lambda \ga 10\mu$m, a changing spectral shape of the 9.7\,$\mu$m silicate
feature, and significant spectral shape changes for $\lambda \la 5\mu$m
emphasizing the need of coeval data to construct the SED.

All variability phases given in this paper refer to a period of $P=660$\,d
and to $\phi=0$ (or 1; maximum light) at JD 2446990 (July 13, 1987) as in
Le Bertre (\cite{LeB93}).
\subsection{Data compilation}
To construct the SED various observations are available, as e.g.\
photometry from the near to the mid-infrared
(Ulrich et al.      \cite{UlrEtal66},
 Wisniewski et al.\ \cite{WisEtal67},
 Hyland et al.\     \cite{HylEtal72},
 Morrison \& Simon  \cite{MorSim73}, 
 Dyck et al.\       \cite{DyckEtal74},
 Strecker \& Ney    \cite{StrNey74},
 Merrill \& Stein   \cite{MerSte76},
 McCarthy et al.\   \cite{McCaEtal77},
 Epchstein et al.\  \cite{EpcEtal80},
 Price \& Murdock    \cite{PriMur83},
 Lockwood           \cite{Lock85},
 Jones et al.\      \cite{JonEtal90},
 Fouque et al.\     \cite{FouEtal92}, 
 Xiong et al.\      \cite{XioEtal94}),
sub-millimeter (Sopka et al.\ \cite{SopEtal85}),
millimeter (Walmsley et al.\ 1991) and
radio (e.g.\ Olnon et al \cite {OlnEtal80}) observations, 
spectrophotometry and spectroscopy
(Lambert et al.\ \cite{LamEtal90},
 Monnier et al.\  \cite{MonEtal98},
 Lancon \& Wood   \cite{LanWoo2000}),  as well as
photometry and spectroscopy obtained by 
IRAS (InfraRed Astronomical Satellite;
      Point Source Cataloque \cite{IRAS85},
      Volk \& Cohen \cite{VolCoh89}) and
ISO (Infrared Space Observatory;
     Sylvester et al.\ \cite{SylEtal99},
     Markwick \& Millar \cite{MarMil2000}). 

%%%%%%%%%%%%%%%%%%%%%%%%%%%%%%%%%%%%%%%%%%%%%%%%%%%%%%%%%%%%%%%%%%%%%%%%%%%%%
%%%% SED photometry and ISO etc.
%%%%%%%%%%%%%%%%%%%%%%%%%%%%%%%%%%%%%%%%%%%%%%%%%%%%%%%%%%%%%%%%%%%%%%%%%%%%%
%\begin{figure}
%\epsfxsize=8.8cm
%\mbox{\epsffile{./plots/obs/aa.sed.all.new.ps}} 
%\caption{
%Spectral energy distribution of \object{CIT\,3}. The symbols refer to
%the observations of Jones et al.\ (1990) [$\phi=0.77$], 
%Sopka et al. (1985) [$\phi=0.89$],  and Walmsley et al.\ (1991)
%[$\phi=0.73$]. Phases were calculated according to Sec.~\ref{SSvar}. 
%Additionally IRAS fluxes (scaled to ISO data) are shown. 
%The line displays the ISO SWS [$\phi=0.22$] and LWS [$\phi=0.49$] 
%spectroscopy of  Sylvester et al.\ (1999). The ISO spectroscopy was 
%adjusted to match the ground-based
%photometry of Jones et al.\ (1990) between $9\mu$m and $18\mu$m.
%}
%\label{Fsedraw} 
%\end{figure}
%%%%%%%%%%%%%%%%%%%%%%%%%%%%%%%%%%%%%%%%%%%%%%%%%%%%%%%%%%%%%%%%%%%%%%%%%%%%%
Figure~\ref{Fbestmod}
displays those observations either taken close to 
or being adustable to the pulsational phase $\phi$ of the 
present speckle measurements covering a
wavelength range from 1\,$\mu$m to 1.3\,mm. 
%%If only magnitudes were given, 
%%the conversion into fluxes was conducted in the photometric
%%system used in the respective publication. 
This concerns the photometry of Jones et al.\ (1990) [$\phi=0.77$], 
the ISO SWS [$\phi=0.22$] and LWS [$\phi=0.49$] spectroscopy of 
Sylvester et al.\ (1999), the 400\,$\mu$m data of Sopka et al. (1985)
[$\phi=0.89$], and the 1.3\,mm data of Walmsley et al.\ (1991)
[$\phi=0.73$]. Phases were calculated according to Sec.~\ref{SSvar}. 
Additionally IRAS fluxes (scaled to ISO data) are shown. 
Since the variability's
brightness amplitude is $\sim$ constant at mid-infrared and longer
wavelenghts, the LWS spectrum taken at minimum light can be adjusted
(factor 1.4, see Sylvester et al. 2000) to meet the SWS spectrum.
Finally the complete ISO spectroscopy was scaled to match the ground-based
photometry of Jones et al.\ (1990) between $9\mu$m and $18\mu$m. 
%%%%%%%%%%%%%% old part %%%%%%%%%%%%%%%%%%%%%%%%%%%%%
%Figure~\ref{Fsedsil} shows the mid-infrared spectrophotometry of
%Monnier et al.\ (1998) for different phases and the interpolated spectrum
%at $\phi=0.75$ suitable for the present phase of the speckle observations
%in comparison with the ISO data. 
%Concerning the silicate feature, absorption is more prominent at
%minimum light,
%the shape of the ISO SWS spectrum (Sylvester et al.\ 2000) at $\phi=0.22$ fits
%better to the miminum than to the maximum light curve.
%%%%%%%%%%%%%%%%%%%%%%%%%%%%%%%%%%%%%%%%%%%%%%%%%%%%%
The inlay of Fig.~\ref{Fbestmod}
shows the mid-infrared spectrum at $\phi=0.75$, interpolated
from the multi-epoch spectrophotometry of Monnier et al.\ (1998),
as suitable for the
present phase of the speckle observations,
in comparison with the ISO data. 
Concerning the silicate feature, absorption is more prominent at minimum light,
accordingly the shape of the ISO SWS spectrum
(Sylvester et al.\ 2000) at $\phi=0.22$ fits better
to the miminum than to the maximum light curve.
%
%%%%%%%%%%%%%%%%%%%%%%%%%%%%%%%%%%%%%%%%%%%%%%%%%%%%%%%%%%%%%%%%%%%%%%%%%%%%%
%%%% Spectrophotometry of Monnier et al. (1998)
%%%%%%%%%%%%%%%%%%%%%%%%%%%%%%%%%%%%%%%%%%%%%%%%%%%%%%%%%%%%%%%%%%%%%%%%%%%%%
%\begin{figure}
%\epsfxsize=8.8cm
%\mbox{\epsffile{./plots/obs/aa.monnier.phases.log2.ps}} 
%\caption{
%Variability-phase dependence of the silicate feature of \object{CIT\,3}.
%The open symbols refer to the spectrophotometry
%of Monnier et al. (1998) taken at phases $\phi$ between
%0.49 and 0.95 except of the 
%circles which give the interpolated spectrum for $\phi = 0.75$, the phase
%of the present speckle observations.
%The filled squares show the photometry of Jones et al. (1990).
%The (lower) solid line corresponds to the ISO SWS spectrum of
%Sylvester et al. (2000), whereas the (upper) dashed one shows the
%ISO spectrum scaled to
%the Jones et al. (1990) photo\-metry matching the $9\mu$m and $18\mu$m flux. 
%}
%\label{Fsedsil} 
%\end{figure}
%%%%%%%%%%%%%%%%%%%%%%%%%%%%%%%%%%%%%%%%%%%%%%%%%%%%%%%%%%%%%%%%%%%%%%%%%%%%%
%
\section{Dust-shell models} \label{Sdusty}
To construct models for the dust shell of \object{CIT\,3} radiative
tranfer calculations were carried out using the code DUSTY
(Ivezi\'c et al. \cite{IveNenEli97}).
The code was expanded for the calculation of synthetic visibilities
as described by Gauger et al.\ (\cite{GauEtal99}).
DUSTY solves the  
radiative transfer problem in spherical symmetry utilizing the
self-similarity and scaling behaviour of IR emission from radiatively
heated dust (Ivezi\'c \&  Elitzur \cite{IveEli97}).
Due to the observed elongation in the $J$-band the assumption of spherical
symmetry appears to be questionable at least on smaller scales. However,
on larger scales and for $\lambda > 2.1\,\mu$m
the dust shell does not deviate significantly from
spherical symmetry and its main properties can be believed to be inferable
from such models in fair approximation. Nevertheless, it should be kept
in mind that the derived dust-shell properties are subject to the geometry
and are, therefore, of tentative character. 

The model construction requires the specification of the central radiation
source, the dust properties (chemistry, grain sizes), the temperature
at the inner dust-shell boundary and the relative
thickness of the dust shell (ratio of outer to inner shell radius),
the  density distribution,  and the total optical depth at a
given reference wavelength.

The models are confronted with the observed SED and the near- and mid-infrared
visibilities. Preference is given to the observations close to the phase
of the present speckle observations ($\phi =0.75$). Concerning the SED,
we take into account the ground-based photometry of Jones et al.\ (1990)
[1.6$\mu$m-18$\mu$m],
the spectrophotometry of Monnier et al.\ (1998) [silicate feature],
the ISO data of Sylvester et al. (1999) [5$\mu$m-100$\mu$m] scaled to the
ground-based photometry, as well as
400$\mu$m (Sopka et al.\ \cite{SopEtal85}) and 1.3mm (Walmsley et al.\ 1991)
data. Interstellar extinction have only minor effects
(Alvarez et al.\ \cite{AlvEtal2000}) and was neglected. For instance,
Xiong et al.\ (\cite{XioEtal94}) gives for the interstellar reddening
$A_{\rm V} = 0.18$\,mag referring to van Herk (\cite{vanHerk65}). 

Previous (spherically symmetric) radiative transfer models were
communicated by, e.g., 
Rowan-Robinson \& Harris (\cite{RowRobHar83}),
Bedijn (\cite{Bed87}),
Schutte \& Tielens (\cite{SchTie89}),
Le Sidaner \& Le\,Bertre (\cite{SidLeB96}), and
Lipman et al.\ (\cite{LipEtal2000}).
The present radiative transfer models
take for the first time a broad range of the SED (from 1\,$\mu$m to 1\,mm)
and visibilities at different wavelengths into account. 

\subsection{Standard single-shell models}
An extensive model grid was calculated for standard uniform-outflow shells,
i.e.\ for density distributions $\sim 1/r^{2}$ within the dust shell.
The central star was approximated by black bodies, however, model atmospheres
(Aringer \& Loidl, private comm.) 
were also taken into account. Due to its spectral type of M9-10,
\object{CIT\,3}'s effective temperature can be estimated to lie in the range
of $\simeq$2500-2700\,K if extrapolated from
the effective temperature scale for
giants given by Perrin et al.\ (\cite{PerEtal98}) (see also
Dyck et al.\ \cite{DyckEtal98}, van Belle et al.\ \cite{VanbEtal99b}).
However, long-period variables can be substantially cooler than giants of the
same spectral type (van Belle et al.\ \cite{VanbEtal99a}), therefore
models were calculated for $T_{\rm eff}=2000$ to 3000\,K.  
Concerning the optical dust properties we considered different sets of
silicate grains (Draine \& Lee \cite{DraLee84},
Ossenkopf et al.\ \cite{OssEtal92}, 
Le Sidaner \& Le\,Bertre \cite{SidLeB93},
David \& Pegourie \cite{DavPeg95}).
In the following sections we will mostly refer to 
the Ossenkopf et al.\ (\cite{OssEtal92}) grains which prove to give good
fits to the silicate feature.

For the grain-size distribution we took into account 
grain-size distributions according to Mathis et al.\
(\cite{MRN77}, hereafter MRN), i.e. $n(a) \sim a^{-3.5}$,
with  0.005\,$\mu {\rm m} \leq a  \leq (0.13$ to  $0.45)$\,$\mu$m   
as well as single-sized grains  with $a=0.05$ to 0.3\,$\mu$m.
Grain-size distributions were accounted for
by the synthetic grain approximation
which first calculates the absorption and scattering effiencies
for each grain and then determines average cross sections
by averaging over the size-distribution. 
The shell thickness was fixed to $Y_{\rm out} = r_{\rm out}/r_{1} = 10^{3}$,
with $r_{\rm out}$ and $r_{1}$ being the outer and inner radius
of the shell, respectively. Larger $Y_{\rm out}$ (e.g.\ $10^{4}$) affect the
SED only slightly beyond 100\,$\mu$m. The model paramaters are
completed by the dust temperature  $T_{1}$
at the shell's inner boundary $r_{1}$, which was varied between 400 and
1500\,K, and the optical depth, $\tau$, at a given reference wavelength,
$\lambda_{\rm ref}$. Refering to 
$\lambda_{\rm ref} =  0.55\,\mu$m, optical depths between 10 and 90 were
considered.

The inspection of the model grid reveals that a simultaneous match of
all observed properties does not appear to be possible within the
considered parameter space. In particular, it turned out that uniform
outflow models cannot provide enough flux in the far-infrared and
simultaneously match the observations at shorter wavelengths. 
This may indicate higher densities in the outer shell regions than given
by the $1/r^{2}$ density distribution, and thus a more
shallow density decline or the existence of regions of enhanced
density due to a previous period of increased mass loss (see, e.g.,
Suh \& Jones \cite{SuhJon97};
Bl\"ocker et al.\ \cite{BloeEtal99},  \cite{BloeEtal01}). 
This is in line 
with the results of Lipman (2000) who found that the ISI visibility data is 
better fitted by a $1/r^{1.5}$ density fall-off than by a uniform outflow.

%%%%%%%%%%%%%%%%%%%%%%%%%%%%%%%%%%%%%%%%%%%%%%%%%%%%%%%%%%%%%%%%%%%%%%%%%%%%
\begin{figure*}
\centering
\epsfxsize=12cm
%%%\mbox{\epsffile{./plots/mod/aua_sfv2_teff_td_tau30_siloc_gdmax025_rm15.v8.ps}} \\[10ex]
\mbox{\epsffile{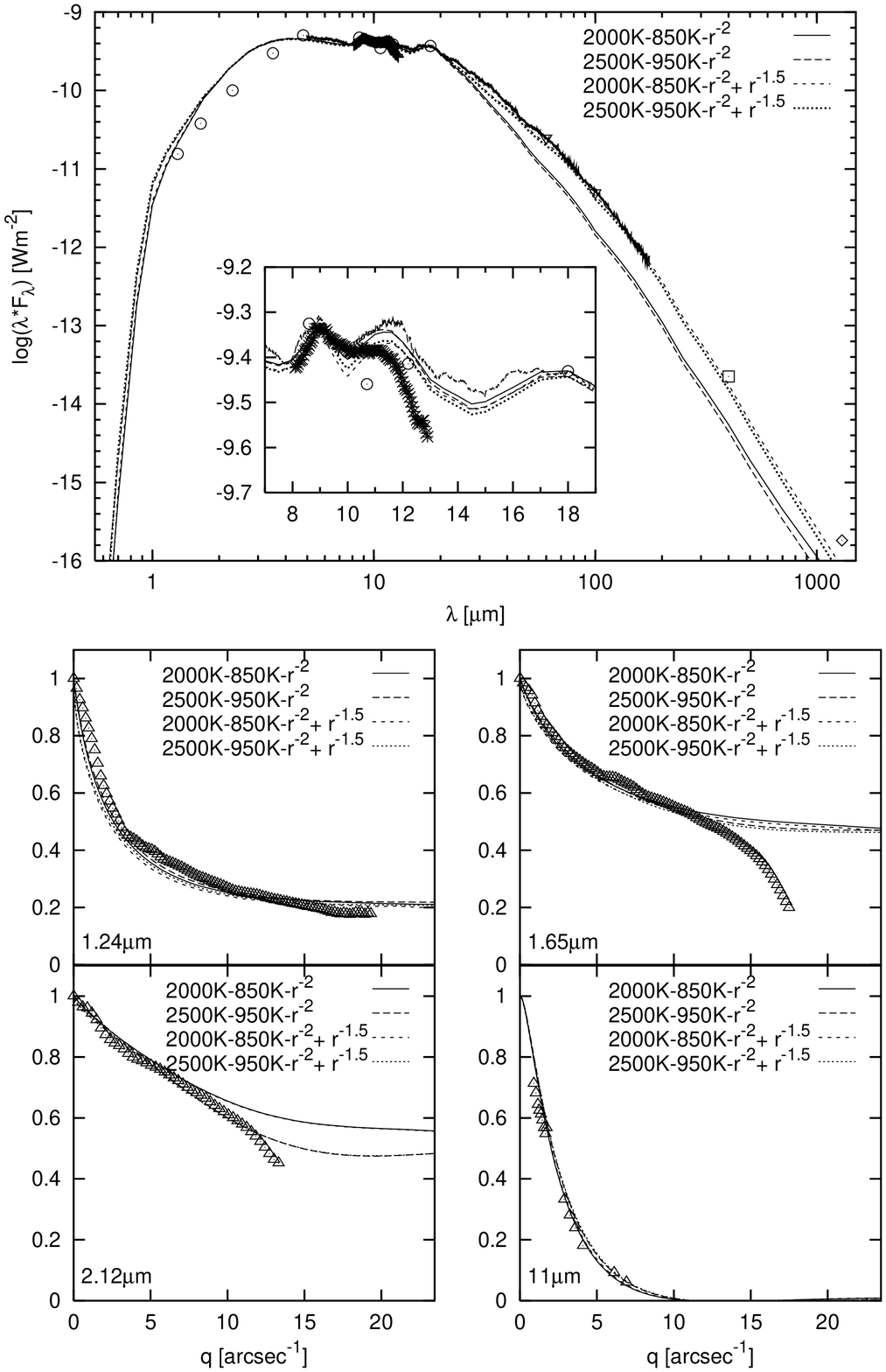}} \\[10ex]
\vspace*{-5mm}
\caption[bestuni]
{Model SED and visibilities at 1.24\,$\mu$m, 1.65\,$\mu$m, 2.12\,$\mu$m, and
11\,$\mu$m for models with
($T_{\rm eff},T_{1}$)=(2000\,K,850\,K) and (2500\,K,950\,K), resp. Both 
uniform outflow models ($\rho \sim 1/r^{2}$; solid and long-dahed lines) 
and models with a flatter density distribution in the outer shell region
($\rho \sim 1/r^{2}$ for $Y \le  20.5$ and  $\rho \sim 1/r^{1.5}$ for 
$Y > 20.5$; short-dashed and dotted lines) are shown. 
The optical depth $\tau_{0.55\mu{\rm m}}$ is 30. 
The calculations are based on a black body, Ossenkopf et al. (1992) silicates,
and an MRN grain size distribution with $a_{\rm max}=0.25\,\mu$m.
The symbols refer to the observations.
SED --- 
{\Large $\circ$}: Jones et al.\ (1990) [$\phi=0.77$]; 
$\Box$: Sopka et al. (1985) [$\phi=0.89$];
$\Diamond $: Walmsley et al.\ (1991) [$\phi=0.73$]; 
$\bigtriangledown$: IRAS fluxes (scaled to ISO data);
{\large $\ast$}:  spectrophotometry of Monnier et al.\ (1998) 
           [interpolated  for $\phi = 0.75$];
%thick line: ISO SWS [$\phi=0.22$] and LWS [$\phi=0.49$] 
%            spectroscopy of  Sylvester et al.\ (1999) 
thick line: ISO spectroscopy of  Sylvester et al.\ (1999) 
            [adjusted to match the 
            Jones et al.\ (1990) photometry at $9\mu$m and $18\mu$m].
The inlay shows the Jones et al.\ (1990) and Monnier et al.\ (1998) data  
together with the ISO spectroscopy of Sylvester et al.\ (1999).
Near-infrared visibilities --- 
$\triangle$: present speckle observations.
Mid-infrared visibility --- $\triangle$: 
ISI observations at  4m, 9.6m and 16m baseline
of Lipman et al.\ (\cite{LipEtal2000}).
\vspace*{-3mm}
}                                      \label{Fbestmod}
\end{figure*}
%%%%%%%%%%%%%%%%%%%%%%%%%%%%%%%%%%%%%%%%%%%%%%%%%%%%%%%%%%%%%%%%%%%%%%%%%%%%%%%
%%%%%% place for SED fits in two-column format
%%%%%%%%%%%%%%%%%%%%%%%%%%%%%%%%%%%%%%%%%%%%%%%%%%%%%%%%%%%%%%%%%%%%%%%%%%%%%%%
The best models found are shown in Fig.~\ref{Fbestmod} and
refer to an optical depth of $\tau_{\rm V}$=30 given 
by the match of the silicate feature.
The grain sizes are strongly contrained by the near-infrared visibilities,
and a grain-size distribution appeared to be better suited than single-sized
grains. Considering a grain-size distribution, the maximum grain-size was
determined to be $a_{\rm max} = 0.25\,\mu$m as in the original
MRN distribution. The temperature at the inner rim of the dust shell was
found to be  in the range of 850-950\,K with the best model at 900\,K for
the best match of visibilities.

The temperature of the
central star is not well contrained by the SED or the visibilities.
For instance, black bodies of 2000\,K and 2500\,K yielded similar overall fits
given the observational and theoretical uncertainties.
However, the lunar occultation observations
of Zappala et al.\ (\cite{ZapEtal74}), yielding a
central-star diameter of 13\,mas and a typical dust-shell
diameter of 65\,mas at $2.2\,\mu$m, seem to favour lower temperatures if one
compares the dimensions of the respective models. For instance, 
adopting $T_{\rm eff}=2000$\,K a bolometric flux of 
$f_{\rm bol}=1 \cdot 10^{-9}$W\,m$^{-2}$ determined by the SED fit 
gives a central-star diameter of $\Theta_{\ast}=13.7$\,mas. 
This leads for $T_{1}=900$\,K, to an inner dust shell diameter of 
$\Theta_{1}=65.8$\,mas corresponding to $r_{1} = 4.8\,R_{\ast}$
in reasonable agreement with the observations.
On the other hand, a star with $T_{\rm eff}=2500$\,K leads  to
a smaller central star of $\Theta_{\ast}=8.8$\,mas and to a dust shell with
$\Theta_{1}=78.2$\,mas and  $r_{1} = 8.9\,R_{\ast}$.
Even though this comparion still lacks corrections due to 
different pulsational phases, it indicates that 
\object{CIT\,3} should be cooler than, say, 2500\,K.
Taking into account the model fits and the above constraint, the best 
suited model is that with  $T_{\rm eff}=2250$\,K giving
 $\Theta_{\ast}=10.9$\,mas and  $\Theta_{1}=73.3$\,mas
($r_{1} = 6.7\,R_{\ast}$).

The phase dependence of the silicate feature can be seen in the inlay of 
Fig.~\ref{Fbestmod}
showing ground-based photometry and spectrophotometry
(Jones et al.\ 1990, Monnier et al.\ 1998) and the ISO data
(Sylvester et al.\ 1999). The difference between the
spectrophotometry (interpolated for $\phi$=0.75) and the ISO data
($\phi$=0.22) becomes most obvious between 10\,$\mu$m and 13\,$\mu$m.
Since the scaled ISO data meets the photometry of Jones et al. (1990)
($\phi$=0.77) both at 9\,$\mu$m and 18\,$\mu$m, the phase dependence of
the spectral shape appeared to be restricted to the above region,
at longer wavelengths flux differences are independent of the wavelength.
The radiative transfer calculations show that for the considered parameters
the spectrophotometry can be matched only in its exact shape on the 
expense of worsening the fits of other observational quantities
as, for instance, the mid-infrared visibility. Additionally, the flux
at 18\,$\mu$m cannot be matched and the flux deficit in the far infrared is
considerably enlarged.
%Therefore, the final model was chosen from those
%matching the SED and the visibilities in generell and being as close as
%possible both to the silicate feature and the 18\,$\mu$m flux. 

The near infrared part of the SED tends to require
somewhat higher optical depths than, for instance, the silicate
feature or the visibilities leading to somewhat higher fluxes than observed.
Regarding the observations this may be attributed to
the fact that the data is for the same pulsational cycle phase
but for different epochs, i.e. to changes of the flux variations during a
pulsational cycle over the last decade.
With respect to the models, the assumption of spherical symmetry has to be
questioned at shorter wavelengthts as obvious in the $J$-band image
of Fig.~\ref{Fima}.
Since the 
spectra of cool giants can deviate considerably from that of a black body
in the optical and near-infrared, we used also
model atmopheres (Aringer \& Loidl, priv comm.,
cf.\ Aringer et al.\ 1997) instead of
simple black bodies for the central source of radiation. The coolest
atmosphere used refer to $T_{\rm eff}=2700$\,K and $\log g = 0$.
Due to the high optical depths, the SED fits remained basically unchanged,
and the visibility curves run only somewhat higher than in the corresponding
black body calculation. For lower effective temperatures as, 
e.g., for $T_{\rm eff}=2250$\,K being appropriate for CIT\,3, 
these differences can be expected to  become larger
due the increasing strength of molecular absorption bands (e.g.\ water vapor) 
which can also be phase dependent (see, e.g., Matsuura et al. 2001).
However, due to the lack of model atmospheres in this regime we 
stayed with the black body approximation. 

We note that,
while the $J$- and $K^{\prime}$-visibilities decline with almost constant slope
at higher spatial frequencies, the $H$-band visibility shows a significant
drop beyond 15 cycles/arcsec. This drop cannot be reproduced by any of the
models. Though still not being significant,
the development of a similar trend seems to be suggested by
the $K^{\prime}$-band data close to the diffraction limit. 
Such a visibility shape can be expected if, for instance, the
central star is partially resolved, which, however, requires a stellar diameter
of the order of 50\,mas and certainly does not apply to CIT\,3.
The reason 
for this drop has to remain open yet but may be related to the asphericity
of CIT\,3. 

\subsection{Different density distributions}
While the above uniform outflow models seem to match the observations 
reasonably well from the near-infrared to the mid-infrared , they fail 
to give sufficiently high fluxes beyond, say, 20\,$\mu$m. 
This indicates, for instance,
 either a flatter density distribution than $\rho \sim 1/r^{2}$ 
as given by the standard uniform outflow model or the existence of 
regions with increased density in the outer shell due to periods of 
higher mass-loss in the past. 
We recalculated the main body of the grid with density distributions 
ranging from $\rho \sim 1/r^{2}$  to $\rho \sim 1/r^{1.5}$ and combinations
of it. It turned out that distributions being generally more shallow 
can match the silicate feature and the mid-infrared visibility but 
fail to fit the near-infrared visibilities. Therefore, 
two-component shells were considered 
with an inner uniform-outflow component and an outer 
component  with a flatter distribution, i.e.\  with 
$\rho \sim 1/r^{x}$ and $x<2$. 
The transition between both density distributions is at the relative radius
$Y_{2} = r_{2}/r_{1}$. 
Models were calculated for $Y_{2}=2.5$ to 30.5.

We also calculated some uniform outflow models with density enhancements or 
superwind components, i.e. with regions in the shell 
where  $\rho \sim A \cdot 1/r^{2}$ with superwind amplitudes $A$ ranging from
10 to 100. However, the spectral index of the far-infrared tail of the SED 
and the visibilities seem to favour the above flat density distribution 
in the outer shell region rather than a superwind component. 

In order to preserve the match of SED and visibilities up to the mid-infrared
the transition to the flatter density distribution should take place 
in sufficient distance from the inner rim of the dust shell. On the 
other hand, in order to provide sufficiently high fluxes in the far-infrared 
the exponent in the density distribution has to sufficiently small. 
The best models found are those with $\rho \sim 1/r^{2}$ for
$Y_{2} \le 20.5$ and $\rho \sim 1/r^{1.5}$ for 
$Y_{2} > 20.5$. The temperature  $T_{1}$ at the inner rim of the dust shell 
ranges again between 850 and 950\,K. Model fits are shown in 
Fig.~\ref{Fbestmod}. Since the  transition  radius $Y_{2} = 20.5$ corresponds 
to a temperature of $T_{2} = 163$\,K,
mostly the outer cool regions of the dust shell are
concerned and the spatial dimension of the dust shell changed only slightly.
For $T_{\rm eff}=2250$\,K and  $T_{1}=900$\,K, 
the inner rim of the dust shell is at $r_{1}/R_{\ast}=6.62$ corresponding 
to an angular diameter of $\Theta_{1}=71.9$\,mas.

These results 
can be compared with the radiative transfer model of Lipman et al.\
(2000). Based on fits of the ISI 11\,$\mu$m visibility data (and the 
silicate feature) they also found 
a $1/r^{1.5}$ density fall-off  to be better suited 
to match the observations. However, an overall $1/r^{1.5}$ density law 
appears to be in conflict with the present 
near-infrared visibilities leading to 
too high densities in the inner parts of the shell. The present modelling
suggests, as mentioned above, the presence of a sufficiently broad 
uniform outflow area in the inner shell region 
before the flatter density distribution takes over. 

%%%%%%%%%%%%%%%%%%%%%%%%%%%%%%%%%%%%%%%%%%%%%%%%%%%%%%%%%%%%%%%%%%%%%%%%%%%%%e
%%%%%%%  Model fits: flux contributions  for two component model
%%%%%%%%%%%%%%%%%%%%%%%%%%%%%%%%%%%%%%%%%%%%%%%%%%%%%%%%%%%%%%%%%%%%%%%%%%%%%%
\begin{figure}  
\centering
\epsfxsize=8.8cm
%%%\mbox{\epsffile{./plots/mod/aua_ffraction_2250_td_900_tau_siloc_gdmax025_20_5_rm15.ps}}
\mbox{\epsffile{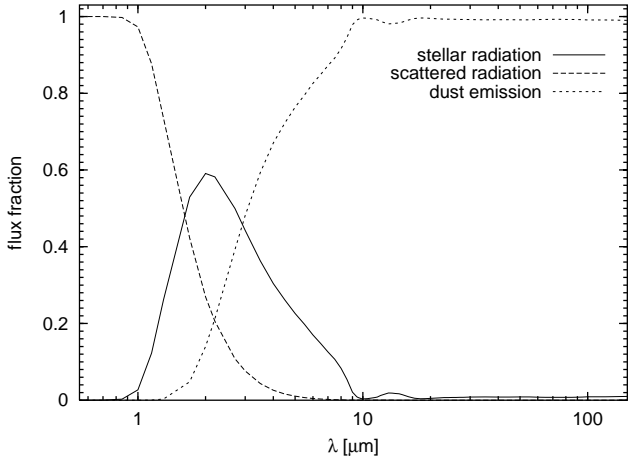}}
\caption[ffraction]
{Fractional contributions of the emerging stellar radiation,
 the scattered radiation, and of the dust emission to the total
 flux as a function of the wavelength for a two-component model.
 The density decreases  $\sim 1/r^{2}$ for $Y_{2} \le 20.5$,
 and   $\sim 1/r^{1.5}$  for  $Y_{2} > 20.5$.
 (see Fig.~\ref{Fbestmod}).
 Model parameters are:
 black body, $T_{\rm eff}=2250$\,K, $T_{1}=900$\,K,
 $\tau_{0.55\mu{\rm m}}=30$, Ossenkopf et al.\ (1992) silicates, and a
 Mathis et al.\ (1977) grain size distribution with $a_{\rm max}=0.25\,\mu$m.
}                                      \label{Fffraction}
\end{figure}
%%%%%%%%%%%%%%%%%%%%%%%%%%%%%%%%%%%%%%%%%%%%%%%%%%%%%%%%%%%%%%%%%%%%%%%%%%%%
%
Fig.~\ref{Fffraction} shows the fractional contributions of the 
emerging stellar radiation, the scattered radiation, 
and of the dust emission to the total flux as a function of the wavelength 
for the above two-component model. 
Whereas the radiation at 11\,$\mu$m is almost completely given by thermal 
dust emission, the radiation in the near-infrared domain is strongly
determined by scattering. The contribution  of scattered light 
amounts to 20.6\%, 42.3\% and 79.2\%
in the $K^{\prime}$-, $H$- and $J$-band, resp., 
while thermal dust emission accounts for 21.2\%, 4.8\% and 0.2\%, resp., 
and the emerging direct stellar light to 58.2\%, 52.9\%, and 20.6\%. resp. 
It demonstrates 
that for the given grain-size distribution 
49.3\%, 89.8\% and 99.7\% of the dust-shell emission is due to scattering. 
We note that these derived attributes of the near-infrared emission
are subject to some uncertainties due to their dependence on the dust
properties and the assumption of spherical symmetry. For instance, 
though the grain-sizes are strongly constrained by the near-infrared 
visibilities, principal ambiguities remain due to, e.g., 
the used optical dust constants depending on shape and chemistry 
of the grains. 
%These uncertainties inherent to radiative transfer calculations can only 
%%be minimized by  given the large number of constraints taken into account 
%% in the present modellin
Since the resolution of the present speckle observations is of the order 
of the inner dust-shell boundary, 
future near-infrared interferometric measurements of higher resolution
can yield further constraints on the models and dust properties, resp.,
by allowing a direct measurement of, e.g.,
the distribution of direct stellar and scattered radiation. 
%
%%%%%%%%%%%%%%%%%%%%%%%%%%%%%%%%%%%%%%%%%%%%%%%%%%%%%%%%%%%%%%%%%%%%%%%%
%%%%%%%  Model fits: Intensity models with 1/r^1.5 in the outer shell
%%%%%%%%%%%%%%%%%%%%%%%%%%%%%%%%%%%%%%%%%%%%%%%%%%%%%%%%%%%%%%%%%%%%%%%%%%%%%%
\begin{figure}
\centering
\epsfxsize=8.8cm
%%%\mbox{\epsffile[36 51 532 774]{./plots/mod/aua_4lam_2250_td_900_tau_siloc_gdmax025_20_5_rm15.ps}}
\mbox{\epsffile[36 51 532 774]{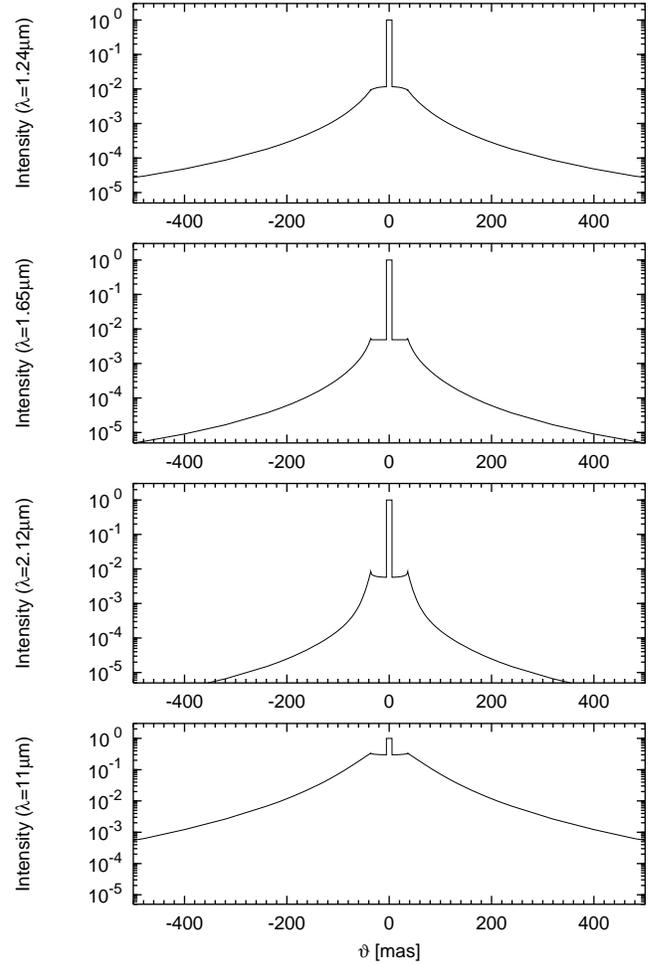}}
\caption[intensity]
{Normalized intensity at $1.24\,\mu$m, $1.65\,\mu$m, $2.12\,\mu$m,
 and $11\mu$m (from top to bottom)
 vs.\ angular displacement $\vartheta$
 for a two component model with a uniform outflow in the inner dust shell 
 region and  $\rho \sim 1/r^{1.5}$  for  $Y_{2} > 20.5$ 
 (see Figs.~\ref{Fbestmod}-\ref{Fffraction}).
 The central peak belongs to the central star with a radius of 5.5\,mas.
 The inner hot rim of the circumstellar shell has a radius of 35.9\,mas,
 the cool far-out component with the flatter density distribution
 is located at a radius of 737\,mas beyond the scale of this figure. 
 In the $K^{\prime}$-band
 the inner rim of the circumstellar shell is noticeable limb-brightened.
 The model refer to  $T_{\rm eff}=2250$\,K, $T_{1}=900$\,K, and
  $\tau (0.55\mu {\rm m}) = 30$.
 The calculations are based on a black body, Ossenkopf et al.\ (1992)
 silicates, and a Mathis et al.\ (1977) grain size distribution with
 $a_{\rm max}=0.25\mu$m.
}                                      \label{Fintensity}
\end{figure}
%%%%%%%%%%%%%%%%%%%%%%%%%%%%%%%%%%%%%%%%%%%%%%%%%%%%%%%%%%%%%%%%%%%%%%%%%%%%%%%
The corresponding spatial intensity distributions at  
$1.24\,\mu$m, $1.65\,\mu$m, $2.12\,\mu$m,  and $11\mu$m are shown 
in Fig.~\ref{Fintensity}.
The optical depths are
$\tau(1.24\,\mu {\rm m})=5.1$,
$\tau(1.65\,\mu {\rm m})=3.1$,
$\tau(2.12\,\mu {\rm m})=1.9$, and
$\tau(11\,\mu {\rm m})=2.2$.
Although exposing an optical depth larger than unity, 
the inner rim of the circumstellar shell is noticeably limb-brightened
in the $K^{\prime}$-band. 

The luminosity of CIT\,3 is 7791\,L$_{\odot}$ if $d=500$\,pc 
and 19945\,L$_{\odot}$ if $d=800$\,pc. The former value would predict 
CIT\,3 to be an AGB star of lower mass whereas the latter indicates a more 
massive AGB star with a core mass larger than, say, 0.8\,M$_{\odot}$. 
Adopting a dust-to-gas ratio of 0.005 and a specific dust density of
3\,g\,cm$^{-3}$,
the present-day mass-loss rate can be determined to be 
$\dot{M} = 1.3 \cdot 10^{-5}$\,M$_{\odot}$/yr if $d=500$\,pc 
and $\dot{M} = 2.1 \cdot 10^{-5}$\,M$_{\odot}$/yr if $d=800$\,pc
which is in line with the
CO observations (see Sect.~\ref{Sintro}).
The smaller exponent of 1.5 in the density distribution beyond $Y_{2}$=20.5
means that the mass-loss rate has decreased with time in the past provided
the outflow velocity kept constant. Adopting $v_{\rm exp}$=20\,km/s, 
the kinematical age of the inner uniform outflow zone can be determined to 
be 87\,yr.

\section{Summary and conclusions} \label{Ssum}
We presented the first near-infrared bispectrum speckle interferometric
observations of  the oxygen-rich, long-period variable AGB star CIT\,3 
in the $J$-, $H$-, and $K^{\prime}$-band. The observations
were taken with the Russian SAO 6m telescope. 
The resolution in the  $J$-band is 88\% of the diffraction-limit (48\,mas), 
in the  $H$-, and $K^{\prime}$-band  the resolution is fully diffraction-limited 
(56\,mas, and 73\,mas, resp.). 

While CIT\,3 appears almost spherically symmetric in the $H$- and $K^{\prime}$-band
it is clearly elongated in the $J$-band along a symmetry axis of position angle
$-28\degr$. Two structures  can be identified: a compact
elliptical core and a fainter north-western fan-like structure.
The eccentricity of the elliptical core, given by the ratio of
minor to major axis, is approximately  $\varepsilon$=123\,mas/154\,mas=0.8.
The full opening angle of the fan amounts to approximately $40\degr$,
i.e.\ it extends from position angle $-8\degr$ to $-48\degr$.

Extensive radiative transfer calculations have been carried out 
and confronted with the observations taking into account 
the spectral energy distribution ranging from 1\,$\mu$m to 1\,mm, 
our near-infrared visibility functions at  1.24\,$\mu$m,
1.65\,$\mu$m and 2.12\,$\mu$m, as well as 
mid-infrared interferometric measurements at $11\,\mu$m.
It has to be noted that these calculations rely on the assumption
of spherical symmetry. Given the observed elongation in the $J$-band, 
this assumption appears to be questionable at least on smaller scales. However,
the main dust properties can be believed to be inferable
from such models in fair approximation
since the deviations from spherical symmetry appear only on smaller scales
and for $\lambda < 2.1\,\mu$m.
Nevertheless, the derived dust-shell properties are subject to the geometry
and are, therefore, of tentative character. It is due to future 
two-dimensional radiative transfer calculations to prove the degree of 
reliance of the presented models. 

The best model found to match all observations refers to a cool central star
with $T_{\rm eff}=2250$\,K which is surrounded by an optically thick 
($\tau (0.55\mu m) = 30$) dust shell. 
The low temperature of the 
central star is constrained by the models but in particular also by 
the lunar occultation observations of Zappala et al.\ (\cite{ZapEtal74})
who derived a stellar diameter of 13\,mas. For the given bolometric flux, this
diameter can only be achieved for low effective temperatures.
The present models give a central-star diameter of $\Theta_{\ast}=10.9$\,mas
being in line with these observations.
Extrapolating the temperature scale of
Perrin et al.\ (\cite{PerEtal98}) based on short-period variables
would predict $T_{\rm eff} \simeq$2500-2700\,K for M9-10 stars.
The lower effective temperature found for CIT\,3 
is consistent with the finding that long-period variables can be 
substantially cooler than giants of the
same spectral type but of shorter period
(van Belle et al.\ \cite{VanbEtal99a}). 

The temperature at the inner 
rim of the dust shell amounts to $T_{1}=900$\,K. 
This is somewhat cooler than 
the usually assumed condensation temperatures for silicate grains of 
1000-1200\,K, but still well within the physical range of dust condensation 
temperatures 
and in reasonable agreement with the results of 
previous radiative transfer models 
(as e.g., 
$T_{1} \sim 850$\,K, Le Sidaner \& Le\,Bertre \cite{SidLeB96}; 
$\sim900$\,K, Bedijn \cite{Bed87} or
$\sim 1000$\,K, Rowan-Robinson \& Harris \cite{RowRobHar83}, 
         Lipman et al.\ \cite{LipEtal2000}).
The inner dust shell boundary is located at 
$r_{1}= 6.6 R_{\ast}$ corresponding 
to an angular diameter of $\Theta_{1}=71.9$\,mas. 
This angular diameter is in accordance with the 2.2\,$\mu$m diameter of 65\,mas
derived by lunar occultation observations (Zappala et al.\ \cite{ZapEtal74})
tracing the hot dust and therefore the very inner parts of the dust shell.

Grains with various optical properties were used, the silicate grains of 
Ossenkopf et al.\ (\cite{OssEtal92}) turned out to be superior to the other 
grain types, for instance in fitting the silicate feature. 
The grain sizes are strongly constrained by the near-infrared visibilities,
and a grain-size distribution proved to be better suited than single-sized
grains. We considered grain-size distribution according to Mathis et al.\
(\cite{MRN77}), i.e. $n(a) \sim a^{-3.5}$,
with  0.005\,$\mu {\rm m} \leq a  \leq a_{\rm max}$   
and determined $a_{\rm max} = 0.25\,\mu$m as in the original
MRN distribution. 

Uniform outflow models, i.e. density distributions with $\rho \sim 1/r^{2}$,
turned out to match the near- to mid-infrared part of the SED and the 
corresponding visibilities but to considerably underestimate the flux beyond
20\,$\mu$m. 
This indicates, for instance,
either a more shallow density distribution than $\rho \sim 1/r^{2}$ 
as given by the standard uniform outflow model or the existence of 
regions with increased density in the outer shell due to periods of 
higher mass-loss in the past. 
The model grid was recalculated for a large variety of accordingly changed
density distributions,
and a two-component model existing of an inner uniform-outflow shell region
($\rho \sim 1/r^{2}$)
and an outer region where the density  declines as $\rho \sim 1/r^{1.5}$
proved to be best suited to match the observations. 
A $\rho \sim 1/r^{1.5}$ density distribution has also been proposed by
Lipman et al.\ (2000) by fits of the $11\,\mu$m visibility. The present 
modelling suggests, however, that the flatter density fall-off should be 
restricted to the outer shell regions which is, for instance, constrained 
by the near-infrared visibilities.
The transition between both density distributions takes place at
$r_{2} = 20.5 r_{1}= 135.7 R_{\ast}$ where the dust-shell temperature has 
dropped to $T_{2} = 163$\,K.
Since these changes concern mostly the very cool regions of the shell,
the other so-far discussed dust-shell properties remain basically unchanged.
Provided the outflow velocity kept constant,
the more shallow density distribution in the outer shell
indicates  that mass-loss had decreased with time in the past of CIT\,3. 
Adopting $v_{\rm exp}=20$\,km/s, the termination of that mass-loss decrease 
and the begin of the uniform-outflow phase took place 87\,yr ago.
The present-day mass-loss rate can be determined to be 
$\dot{M} = 1.3 \cdot 10^{-5}$\,M$_{\odot}$/yr if $d=500$\,pc 
and $\dot{M} = 2.1 \cdot 10^{-5}$\,M$_{\odot}$/yr if $d=800$\,pc
being in agreement  with the CO observations, and 
the corresponding stellar luminosity is 7791\,L$_{\odot}$ 
and 19945\,L$_{\odot}$, resp.

CIT\,3 proved to be among the most interesting far-evolved 
AGB stars due to its infrared properties. Moreover, 
the aspherical appearence of its dust shell in the $J$-band 
puts it in one line with the few AGB stars known to expose near-infrared 
asphericities in their dust shells.
%This includes, for instance, 
%\object{AFGL\,2290} (Gauger et al.\ \cite{GauEtal99})
%being also oxygen-rich, or the carbon stars
%CIT\,6 (Monnier et al.\ 2000) and
%\object{IRC+10216} (see e.g.\ Osterbart et al.\ 2000). 
The development of such asphericities close to the central
star suggests that it is in the very end of the AGB evolution
or even in transition to the proto-planetary nebula phase where 
most objects are observed in axi\-symmetric geometry (Olofsson 1996).
However, in contrast to other objects, CIT\,3 shows these
deviations from spherical symmetry still only in the $J$-band,
which is almost completely dominated by scattered light. This suggests that
CIT\,3 had just started to form aspherical structures and is in this regard 
still in the beginning of its very final AGB phase. If so, CIT\,3 is one 
of the earliest representatives of this dust-shell transformation phase
known so far.  

\begin{acknowledgements}
The observations were made with  the SAO 6\,m telescope, operated by the
Special Astrophysical Observatory, Russia.
We thank Roger Sylvester for providing the ISO spectral energy distribution,
and Rita Loidl and Bernhard Aringer for sending us their model atmospheres.
The radiative-transfer calculations are based on 
the code DUSTY developed by \v{Z}.\ Ivezi\'c, M.\ Nenkova and M. Elitzur.  
This research has made use of the SIMBAD database, operated by CDS in
Strasbourg, and of the NASA's Astrophysics Data System.
\end{acknowledgements}

\end{document}